# RF Injection Locking of THz Metasurface Quantum-Cascade VECSEL

Yu Wu,[1*] Christopher A. Curwen,[2] Mohammad Shahili,[1] John L. Reno,[3] Benjamin S. Williams[1]

[1]Department of Electrical and Computer Engineering, University of California Los Angeles, Los Angeles, California 90095, USA
[2]Jet Propulsion Laboratory, California Institute of Technology, Pasadena, California 91109, USA
[3]Sandia National Laboratories, Center of Integrated Nanotechnologies, MS 1303, Albuquerque, New Mexico 87185, USA
*Corresponding author: ywu17@ucla.edu

**Abstract: RF injection locking and spectral broadening of a terahertz (THz) quantum-cascade vertical-external-cavity surface-emitting laser (QC-VECSEL) is demonstrated. An intra-cryostat VECSEL focusing cavity design is used to enable continuous-wave lasing with a cavity length over 30 mm which corresponds to a round-trip frequency near 5 GHz. Strong RF current modulation is injected to the QC-metasurface electrical bias to pull and lock the round-trip frequency. The injection locking range at various RF injection powers is recorded and compared with the injection locking theory. Moreover, the lasing spectrum broadens from 14 GHz in free-running mode to a maximum spectral width around 100 GHz with 20 dBm of injected RF power. This experimental setup is suitable for further exploration of active mode-locking and picosecond pulse generation in THz QC-VECSELs.**

## 1. Introduction

The terahertz (THz) spectral region has a need for high-resolution, high-speed spectroscopy techniques, as many gas phase polar molecular species have strong characteristic rotational lines there. Examples include applications in industrial and environmental monitoring,[1,2] chemical detection and identification,[3,4] combustion diagnostics.[5] The quantum cascade (QC) laser is well suited for spectroscopic applications as it has been demonstrated as a compact, electrically pumped semiconductor source which gives high power, broadband, coherent THz radiation.[6–8] Its inherently high optical nonlinearity induces self-phase locking through four-wave mixing, which promotes the generation of spontaneous frequency combs; these have been demonstrated in waveguide-based Fabry-Pérot[9–11] and ring QC-lasers.[12,13] Based on that, THz dual-comb spectroscopy has been demonstrated, surpassing the precision and speed of traditional Fourier spectrometers by several orders of magnitude.[14–17]

In separate experiments, THz quantum-cascade lasers have recently been implemented in the vertical-external-cavity surface-emitting laser (VECSEL) architecture, which exhibits watt-level output power, near diffraction-limited beam quality, and ~20% continuous fractional single-mode tuning.[18–20] The key component of a QC-VECSEL is an amplifying reflectarray metasurface of metal-metal waveguide antennas that are loaded with QC-gain material. It is further paired with a partially transmissive output coupler to form the laser cavity. In contrast to ridge-waveguide QC-lasers, experiments have shown that QC-VECSELs tend to operate in single-mode regime despite having large gain bandwidths. For example, we have developed an intra-cryostat focusing VECSEL cavity to reduce the intra-cavity diffraction loss and enable continuous-wave (CW) lasing at 3.4 THz with a cavity length of ~30 mm.[21] Even though the gain bandwidth of the metasurface used was at least 100 times larger than the free spectral range, only a single lasing mode was observed. This is mainly due to a lack of spatial hole burning within the QC-VECSEL metasurface which suppresses multi-mode instabilities,[22] although it is perhaps compounded by the fact that no effort towards dispersion engineering has yet been attempted. Despite these challenges, there is strong interest in achieving active mode-locked QCLs and frequency combs within the QC-VECSEL architecture.

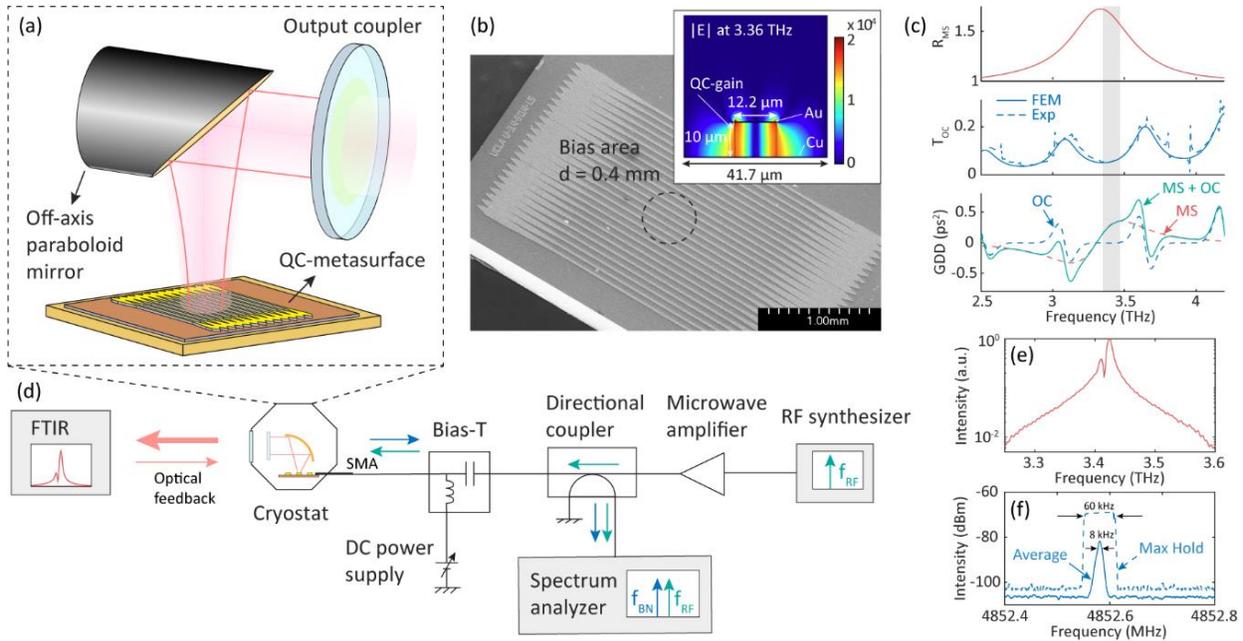

**Figure 1.** (a) Schematic of the QC-VECSEL based on an intra-cryostat focusing cavity design. (b) Scanning electron microscopy image of the fabricated QC-metasurface. The inset shows the dimension and E-field distribution in a single ridge antenna. (c) FEM simulated active metasurface reflectance, output coupler transmittance and GDD contributed by the two components. Shaded area indicates the frequency range where lasing is observed. (d) Schematic of experimental setup for RF injection locking. THz lasing spectrum (e) and beat note spectra (f) of the free-running QC-device in existence of optical feedback are collected at a DC bias of 0.235 mA, where optical feedback is provided by the moving FTIR mirror. The beat note spectra are measured in both Average (solid) and Max Hold modes (dashed) with a RBW of 2 kHz.

Radiofrequency (RF) current modulation of QC-lasers has been demonstrated to promote the generation of sidemodes; by injecting a RF signal near the cavity round-trip frequency, the generated sidemodes will lock existing adjacent free-running lasing modes or seed new ones, which allows for the stabilization and tuning of frequency comb states.[23–26] RF modulation and injection locking is also an important mechanism for active mode-locking in QC-lasers; pulses as narrow as 4-5 ps have been reported.[27,28]

In this article, we demonstrate the emergence of spectral broadening and multimoding in THz QC-VECSEL as we inject strong RF current modulation to the QC-metasurface at a frequency close to the cavity round-trip frequency; at the same time, round-trip frequency pulling and locking to the injected RF signal is observed. The lasing bandwidth and injection locking range increase monotonically with the injected RF power. Lasing modes spanning ~100 GHz are demonstrated under an injected RF power of 20 dBm at 4852.7 MHz, along with a locking range ~5 MHz.

## 2. Sample and experimental setup

The QC-VECSEL used for all measurements is based on an intra-cryostat focusing cavity design, as sketched in Figure 1(a). An off-axis paraboloid (OAP) mirror with a focal length of 12.7 mm is introduced into the VECSEL cavity to reduce the intra-cavity diffraction loss and enable CW lasing using small metasurfaces in long lasing cavities.[21] The QC-metasurface used in this paper is the same one as reported in ref [21], with a small bias area of diameter $d = 0.4$ mm for reduced injection current. It is designed to be resonant at 3.3 THz and consists of an array of ridges of width 12.2 µm repeated with a period of 41.7 µm (Figure 1(b)). An output coupler with $R_{OC} \approx$ 95% is used in pair with the metasurface to form a laser cavity. Both two components are dispersive and contribute to the group delay dispersion (GDD) over one round trip - it exhibits a maximum value exceeding 0.35 ps$^2$ in the frequency range where lasing occurs. The simulated spectral response of the metasurface and the output coupler are plotted in Figure 1(c) based on full-wave 2D finite-element (FEM) electromagnetic reflectance simulation (Ansys HFSS). Detailed information of the active region design and simulation parameters can be found in the Supporting Information.

The experimental setup for RF injection locking is depicted in Figure 1(d). All the measurements were performed in vacuum at a temperature of 77 K. We note that formable semi-rigid coaxial cable is used within the cryostat up to the chip carrier package (see Supporting Information). Due to impedance mismatch between the

50Ω SMA port and the QC-device, the spectrum analyzer collects not only the generated beat note from the QC-device $f_{BN}$ (blue arrow), but also the RF injection signal reflected at the interface of SMA/QC-device $f_{RF}$ (green arrow).

In free-running case, although only single-mode lasing was observed using the same QC-device in ref [21], here we note that the existence of optical feedback can induce multi-mode operation.[29–31] Due to the temporally varying small feedback from the scanning Fourier-transform infrared spectrometer (FTIR, Nicolet 8700) mirrors, we observed at least two lasing modes separated by ~14 GHz in the emission spectrum (Figure 1(e)). Additional low intensity sidemodes may also exist but cannot be resolved by the limited FTIR resolution of 7.5 GHz. This phenomenon is similar as that observed in ref [30], where additional lasing modes are observed in a Mid-IR QC-laser under tilted optical feedback. Through nonlinear mixing of the free-running lasing modes, a weak electrical beat note signal is observed. It is collected using the spectrum analyzer (Agilent N9020a) both in Average mode and in Max Hold mode over 30 seconds, which indicates a round-trip frequency $f_{BN} \approx$ 4852.7 MHz and an equivalent cavity length around 31 mm (Figure 1(f)). A narrow 8-kHz -3dB linewidth in Average mode and 60-kHz linewidth in Max Hold mode of the intermodal beat note is on the same order as the free-running linewidth of the single THz lasing mode measured in [21]; it is likely contributed by only two (or a few) lasing modes. Furthermore, recent study of optical feedback has highlighted its effect on THz QC-lasers combs.[32–34] Here, we experimentally demonstrate that optical feedback plays an important role in determining not only the free-running beat note frequency but also the injection locking range (see Supporting Information).

## 3. RF injection locking

With the knowledge of an accurate round-trip frequency, we are able to systematically study the modulation-dependent behavior of this QC-device. We swept the RF modulation frequency around the round-trip frequency at various modulation powers from -20 dBm to 20 dBm. All RF powers indicated in this paper refer to the nominal output level of the RF synthesizer (Hewlett-Packard 83650B) or after a 20 dBm amplifier (Hewlett-Packard 8349B). The DC bias is fixed at a current of 0.235 mA ($\approx 1.17 \times I_{th}$), and the THz emission spectra as well as intermodal beat note are collected and plotted in Figure 2.

At the lowest power level of -20 dBm, the spectral map in Figure 2(a) clearly shows that the beat note is pulled toward the injection signal and finally locked. A locking range of 30 kHz is demonstrated which increases with respect to the RF injection power. Starting from an RF power of -2.5 dBm, injection locking occurs before the beat note is fully pulled to meet the injected signal $f_{RF}$ (Figure 2(b)); at the same time, lasing bandwidth broadening is observed in the THz emission spectra. This spectral broadening increases with respect to RF power as shown in Figure 2(d). The maximum RF injection power used in this measurement is 20 dBm limited by the max allowable power of the bias-Tee. THz emission and RF beat note spectral maps in this case are plotted in Figure 2(e-f). The maximum spectral broadening occurs at $f_{RF}$ = 4852.7 MHz with lasing modes spanning around 100 GHz (Figure 2(g)). However, due to the limited FTIR resolution of 7.5 GHz, we were not able to spectrally resolve individual lasing modes. The corresponding power and voltage vs. current (P-I-V) curves are plotted in Figure 2(h) (solid curve). A maximum output power around 10 mW was collected using a pyroelectric detector (GentecEO). Compared with the P-I characteristic in the free-running case (dashed curve), the output power, as well as the lasing threshold, is slightly lower. It is noticed from Figure 2(e) that the symmetry of the lasing spectrum is highest at $f_{RF}$ = 4852.7 MHz, where the maximum bandwidth is observed with relative low THz output power obtained from the P-I curve. At injection frequencies above/below this value, the optical power increases – still smaller than that in free-running case – and concentrates toward lower/higher portion of the spectrum. This phenomenon is similar as that reported in ref [26] and a possible explanation can be found in ref [35] due to phase mismatch between the modulation period and group round-trip time. In Figure 2(f), although there is no beat note pulling observed, it is notable that the emission spectrum undergoes distinct change as the beat note disappears (pointed out by red arrows) – it is believed that this is a signature of injection locking and occurs in our measurements under different RF powers.

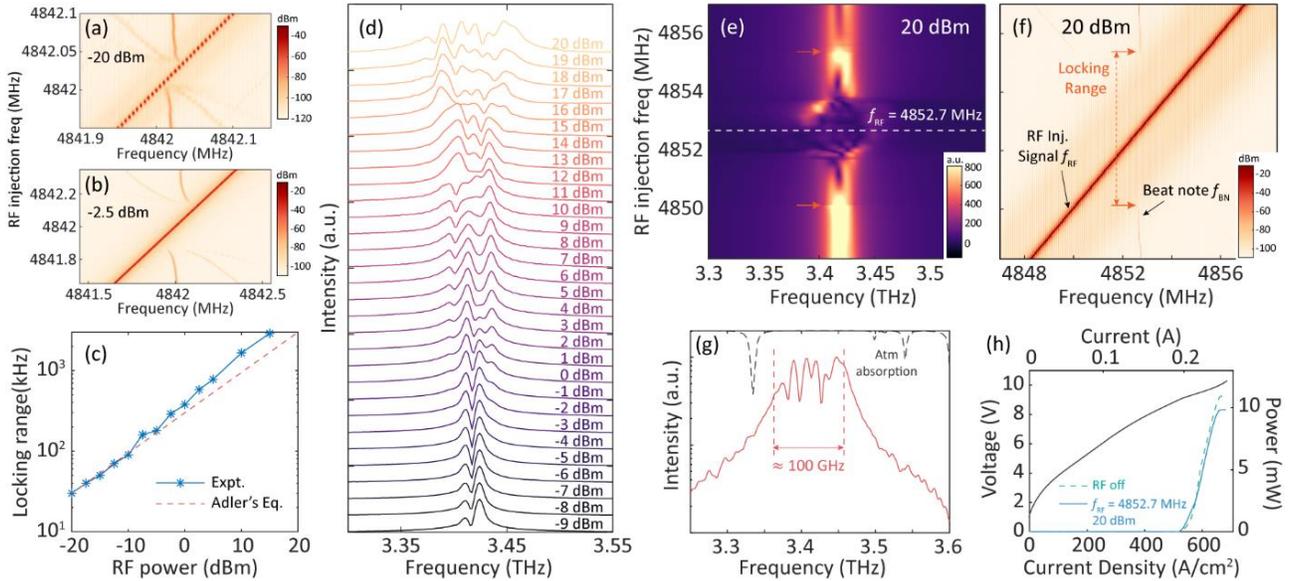

**Figure 2.** Beat note spectral map under constant RF injection power of -20 dBm (a) and -2.5 dBm (b) with RF modulation frequency sweeping around the round-trip frequency. (c) Experimental injection locking range at different RF injection powers (blue stars), following a 0.5-slope dependence in log–log scale (red dashed line). The free-running beat note frequency was shifted from ~ 4842 MHz in (a-c) to ~ 4853 MHz in (d-h) as the movement of cryostat changes the amount of optical feedback. (d) THz lasing spectra at increasing RF power when $f_{RF}$ is fixed at 4852.7 MHz. (e) Lasing spectral and (f) beat note maps of the device under constant RF injection power of 20 dBm. The estimated locking range is pointed out by the red arrows. The maximum spectral broadening occurs at $f_{RF}$ = 4852.7 MHz (white dashed line) and the THz lasing spectrum and P-I-V curves in this case are plotted in (g-h).

The experimental locking range at various RF injection powers is plotted in Figure 2(c). To analyze the phenomenon of RF injection locking, Adler's equation is commonly used with a locking bandwidth given by:[23,36]

$$\Delta \nu = \frac{2\nu_0}{Q}\sqrt{\frac{P_{inj}}{P_0}}, \qquad (1)$$

where $Q$ is the cold-cavity quality factor, $\nu_0$ and $P_0$ are the frequency and power of a free-running longitudinal mode, while $P_{inj}$ is the power of the injected sideband induced by RF injection. Adler's equation indicates a square root dependence of the locking bandwidth on the RF power and fits our experimental results well at low RF powers (red dashed line). However, our experimental locking range deviates from Adler's equation towards higher values under strong RF modulation. This may indicate the limitation of Adler's equation in explaining RF injection locking especially in the case when multiple new lasing modes are excited at RF powers > −2.5 dBm. Adler's equation assumes a weak injection signal where amplitude perturbation induced by the injection signal is not considered; a more rigorous derivation of the locking range is therefore needed.

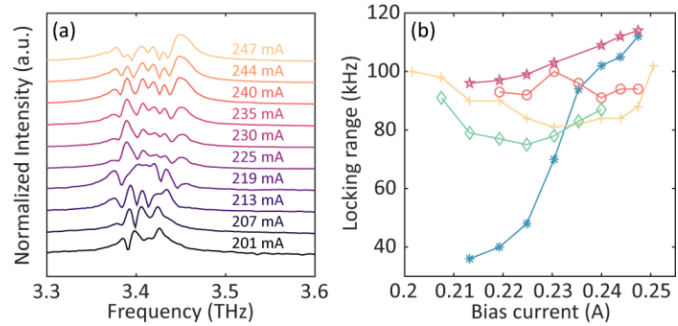

**Figure 3.** (a) THz lasing spectra at various biases when RF signal at 4852.7 MHz is injected into the QC-device, the RF power used is 20 dBm for significant spectral broadening. (b) Injection locking range as a function of bias current under -15 dBm RF power in the cases of different length/strength/angle of optical feedback.

Moreover, we studied the behavior of this QC-device at various DC biases ranging from the lasing threshold to near the NDR point. Figure 3(a) shows the lasing spectra under 20 dBm RF injection at a frequency of $f_{RF}$ = 4852.7 MHz. Significant spectral broadening is observed at all the applied biases, and the lasing bandwidth increases only slightly with respect to the bias current as more modes are brought above the lasing threshold.

As a next step, the effects of device bias on the injection locking bandwidth were investigated. We swept the RF modulation frequency around the round-trip frequency at a fixed injection power of -15 dBm and measured the injection locking bandwidth at various biases. Small injection power was used so the locking range can be more clearly observed. Additionally, we repeated such bias sweeps while providing different magnitude, phase, and angle of feedback light from an external mirror, the corresponding locking ranges are indicated by different colored curves in Figure 3(b). Our experimental observation reveals that the relationship between locking range and device bias is related to the condition of optical feedback, i.e. feedback length (phase), strength and tilted angle. This is significantly different from previous demonstrations using ridge-waveguide QC-lasers, where the locking range became smaller with increasing bias.[23,37] In our system, we could make a simple assumption that there are two free-running modes, where mode $\omega_1$ is induced by optical feedback around the main lasing peak $\omega_0$ and is locked by the RF-excited sideband of the latter. The ratio of $P_{inj}/P_0$ in Adler's equation can be estimated as the ratio of the free-running power of mode $\omega_0$ and that of mode $\omega_1$ as the injected RF power is fixed, which determines the injection locking range. How the locking range changes is therefore determined by how the relative power of two lasing modes develops with respect to bias. Unfortunately, this is not able to be observed experimentally limited by the resolution of our FTIR. In theory, the spectral characteristics of the device versus applied bias is expected to be affected by the changes of threshold gain induced by optical feedback and the alignment of compound-cavity modes formed in the external cavity with respect to gain, which is related to not only the length and strength of optical feedback, but also the tilt angle of external mirror.[31] To fully understand this phenomenon, a theoretical study of laser dynamics and instabilities of QC-VECSELs under optical feedback and systematic experiments of the RF-injected system with well-controlled, adjustable optical feedback will be needed and are beyond the scope of this paper.

## 4. Discussion and conclusion

The injection locking range obtained in this paper is considerably smaller compared with those demonstrated in RF injection-locked Fabry-Pérot waveguide QC-lasers at same level of RF power.[23,25] One of the reasons is that QC-VECSELs have higher quality factors compared with ridge waveguide QC-lasers. Our VECSEL has a 31 mm-long external cavity and low loss from the ~95% reflectance output coupler; using a coupled-cavity model we estimate a cold-cavity linewidth of

$v_0/Q \approx 70$ MHz. This is around 300 times smaller than a value of 25 GHz estimated in ref [23]. In addition, intrinsic and technical issues with our QC-VECSEL setup result in a low efficiency of RF power transfer at ~4.8 GHz from the synthesizer to the QC-metasurface bias terminal. First, due to parasitic capacitances contributed by unbiased regions, the QC-metasurface itself exhibits a larger RC time-constant compared with a narrow ridge waveguide. Second, the electrical packaging has not been optimized for RF operation, where wire bonds and wire bonding pads contribute parasitic inductance and capacitance respectively. Consequently, there is a huge impedance mismatch between the 50Ω SMA port and the QC-device, the resulting transmittance of RF signal through the SMA/QC-package boundary is simulated to be ~4% at a target frequency of 4.8 GHz (see Supporting Information), only part of which will be applied to modulate the gain material. To make things worse, an additional ~8 dB RF attenuation has been characterized accounting for losses through cables and directional coupler from the synthesizer to the SMA connector. In contrast to other demonstrations of ridge waveguide QC-lasers using RF coplanar probes,[23,38] RF launchers,[39] or custom high-frequency PCB mounts[28] to achieve modulation of QC-lasers up to 35 GHz, microwave rectification technique indicates a significant roll-off at frequency higher than 3 GHz in our QC-device (see Supporting Information).

In conclusion, we demonstrate RF injection locking in a THz QC-VECSEL based on intra-cryostat focusing cavity design. Round-trip frequency pulling and locking against an RF injection signal is observed. Furthermore, the RF amplitude modulation leads to broadening of the lasing spectrum up to a spectral width of 100 GHz. This is particularly notable, as multi-mode lasing in QC-VECSELs has been extremely difficult to achieve due to the lack of spatial hole burning within the metasurface; before now at most 9 lasing modes had been observed.[22] There are several obvious avenues for improvement. First, RF attenuation and impedance mismatch severely limits the modulation efficiency, and strong RF reflections impede the detection of the electrical beat note signal using a spectrum analyzer. This can be improved by optimizing the electrical packaging of the QC-device, i.e. reducing the capacitance and inductance portion of the equivalent circuit by 1) redesigning the QC-metasurface with reduced unbiased area and an improved RF feed structure; 2) replacing the electrical contact pad with a well-designed PCB 50Ω transmission line feed up to the edge of the metasurface chip with minimal wire bond length. Second, we note that no particular effort to provide dispersion compensation has been attempted here; further engineering of GDD within the QC-VECSEL cavity may be needed to increase the lasing across the entire ~1 THz gain bandwidth. Finally, given measurements of ridge-waveguide THz QC-lasers under strong RF modulation, it is quite likely that this device is generating short pulses in an active mode-locking regime.[27,28] Further characterization techniques such as shifted-wave interference Fourier-transform spectroscopy (SWIFTs)[10,40,41] or asynchronous electro-optical sampling will be needed to recover the time-domain structure of the field.[42,43]

**Supporting Information**
Supporting Information is available from the author.

**Acknowledgments**
The authors thank David Burghoff, Andres Forrer, Giacomo Scalari, and Stefano Barbieri for valuable conversations. Microfabrication was performed at the UCLA Nanoelectronics Research Facility, wire bonding was performed at the UCLA Center for High Frequency Electronics. This work was performed, in part, at the Center for Integrated Nanotechnologies, an Office of Science User Facility operated for the U.S. Department of Energy (DOE) Office of Science. Sandia National Laboratories is a multimission laboratory managed and operated by National Technology and Engineering Solution of Sandia, LLC., a wholly owned subsidiary of Honeywell International, Inc., for the U.S. Department of Energy's National Nuclear Security Administration under contract DE-NA-0003525. Partial funding was provided by the National Science Foundation (2041165), and the National Aeronautics and Space Administration (80NSSC19K0700).

**References**
[1]  D. M. Mittleman, R. H. Jacobsen, R. Neelamani, R. G. Baraniuk, M. C. Nuss, *Appl. Phys. B Lasers Opt.* **1998**, *67*, 379.
[2]  A. Cuisset, F. Hindle, G. Mouret, R. Bocquet, J. Bruckhuisen, J. Decker, A. Pienkina, C. Bray, É. Fertein, V. Boudon, *Appl. Sci.* **2021**, *11*, 1229.
[3]  H. Zhong, A. Redo-Sanchez, X.-C. Zhang, *Opt. Express* **2006**, *14*, 9130.
[4]  I. R. Medvedev, C. F. Neese, G. M. Plummer, F. C. De Lucia, *Opt. Lett.* **2010**, *35*, 1533.
[5]  M. Araki, K. Matsuyama, *Curr. Appl. Phys.* **2022**, *36*, 83.
[6]  R. Kohler, A. Tredicucci, F. Beltram, H. E. Beere, E. H. Linfield, A. G. Davies, D. A. Ritchie, R. C. Iotti, F. Rossi, *Nature* **2002**, *417*, 156.
[7]  L. H. Li, L. Chen, J. R. Freeman, M. Salih, P. Dean, A. G. Davies, E. H. Linfield, *Electron. Lett.* **2017**, *53*, 799.
[8]  D. Turčinková, G. Scalari, F. Castellano, M. I. Amanti, M. Beck, J. Faist, *Appl. Phys. Lett.* **2011**, *99*, 191104.
[9]  M. Rösch, G. Scalari, M. Beck, J. Faist, *Nat. Photonics* **2014**, *9*, 42.
[10] D. Burghoff, T. Y. Kao, N. Han, C. W. I. Chan, X. Cai, Y. Yang, D. J. Hayton, J. R. Gao, J. L. Reno, Q. Hu, *Nat. Photonics* **2014**, *8*, 462.
[11] A. Forrer, Y. Wang, M. Beck, A. Belyanin, J. Faist, G. Scalari, *Appl. Phys. Lett.* **2021**, *118*, 131112.
[12] M. Piccardo, B. Schwarz, D. Kazakov, M. Beiser, N. Opačak, Y. Wang, S. Jha, J. Hillbrand, M. Tamagnone, W. T. Chen, A. Y. Zhu, L. L. Columbo, A. Belyanin, F. Capasso, *Nature* **2020**, *582*, 360.
[13] M. Jaidl, N. Opačak, M. A. Kainz, S. Schönhuber, D. Theiner, B. Limbacher, M. Beiser, M. Giparakis, A. M. Andrews, G. Strasser, B. Schwarz, J. Darmo, K. Unterrainer, *Optica* **2021**, *8*, 780.

[14] M. Rösch, G. Scalari, G. Villares, L. Bosco, M. Beck, J. Faist, *Appl. Phys. Lett.* **2016**, *108*, 171104.
[15] L. A. Sterczewski, J. Westberg, Y. Yang, D. Burghoff, J. Reno, Q. Hu, G. Wysocki, *Optica* **2019**, *6*, 766.
[16] H. Li, Z. Li, W. Wan, K. Zhou, X. Liao, S. Yang, C. Wang, J. C. Cao, H. Zeng, *ACS Photonics* **2020**, *7*, 49.
[17] L. Consolino, M. Nafa, M. De Regis, F. Cappelli, K. Garrasi, F. P. Mezzapesa, L. Li, A. G. Davies, E. H. Linfield, M. S. Vitiello, S. Bartalini, P. De Natale, *Commun. Phys.* **2020**, *3*, 1.
[18] L. Xu, C. A. Curwen, P. W. C. Hon, Q.-S. Chen, T. Itoh, B. S. Williams, *Appl. Phys. Lett.* **2015**, *107*, 221105.
[19] C. A. Curwen, J. L. Reno, B. S. Williams, *Appl. Phys. Lett.* **2018**, *113*, 011104.
[20] C. A. Curwen, J. L. Reno, B. S. Williams, *Nat. Photonics* **2019**, *13*, 855.
[21] Y. Wu, C. Curwen, J. L. Reno, B. Williams, *Appl. Phys. Lett.* **2022**, *121*, 191106.
[22] Y. Wu, S. Addamane, J. L. Reno, B. S. Williams, *Appl. Phys. Lett.* **2021**, *119*, 111103.
[23] P. Gellie, S. Barbieri, J. F. Lampin, P. Filloux, C. Manquest, C. Sirtori, I. Sagnes, S. P. Khanna, E. H. Linfield, H. E. Beere, D. A. Ritchie, *Opt. Express* **2010**, *18*, 20799.
[24] H. Li, P. Laffaille, D. Gacemi, M. Apfel, C. Sirtori, J. Leonardon, G. Santarelli, M. Rösch, G. Scalari, M. Beck, J. Faist, W. Hänsel, R. Holzwarth, S. Barbieri, *Opt. Express* **2015**, *23*, 33270.
[25] A. Forrer, L. Bosco, M. Beck, J. Faist, G. Scalari, *Photonics* **2020**, *7*, 9.
[26] B. Schneider, F. Kapsalidis, M. Bertrand, M. Singleton, J. Hillbrand, M. Beck, J. Faist, *Laser Photonics Rev.* **2021**, *15*, 2100242.
[27] A. Mottaghizadeh, D. Gacemi, P. Laffaille, H. Li, M. Amanti, C. Sirtori, G. Santarelli, W. Hänsel, R. Holzwart, L. H. Li, E. H. Linfield, S. Barbieri, *Optica* **2017**, *4*, 168.
[28] U. Senica, A. Forrer, T. Olariu, P. Micheletti, S. Cibella, G. Torrioli, M. Beck, Jerome Faist, G. Scalari, *arXiv:2207.06737 [physics.optics]*.
[29] X. Qi, K. Bertling, T. Taimre, G. Agnew, Y. L. Lim, *Phys. Rev. A* **2021**, *103*, 033504.
[30] X. G. Wang, B. Bin Zhao, Y. Deng, V. Kovanis, C. Wang, *Phys. Rev. A* **2021**, *103*, 23528.
[31] D. S. Seo, J. D. Park, J. G. Mcinerney, M. Osinski, *IEEE J. Quantum Electron.* **1989**, *25*, 2229.
[32] M. Wienold, B. Röben, L. Schrottke, H. T. Grahn, *Opt. Express* **2014**, *22*, 30410.
[33] X. Liao, X. Wang, K. Zhou, W. Guan, Z. Li, X. Ma, C. Wang, J. C. Cao, C. Wang, H. Li, *Opt. Express* **2022**, *30*, 35937.
[34] M. Piccardo, P. Chevalier, T. S. Mansuripur, D. Kazakov, Y. Wang, N. A. Rubin, L. Meadowcroft, A. Belyanin, F. Capasso, *Opt. Express* **2018**, *26*, 9464.
[35] Y. Wang, A. Belyanin, *Opt. Express* **2015**, *23*, 4173.
[36] M. R. St-Jean, M. I. Amanti, A. Bernard, A. Calvar, A. Bismuto, E. Gini, M. Beck, J. Faist, H.C.Liu, Carlo Sirtori, *Laser Photon. Rev.* **2014**, *8*, 443.
[37] F. P. Mezzapesa, K. Garrasi, J. Schmidt, L. Salemi, V. Pistore, L. Li, A. G. Davies, E. H. Linfield, M. Riesch, C. Jirauschek, T. Carey, F. Torrisi, A. C. Ferrari, M. S. Vitiello, *ACS Photonics* **2020**, *7*, 3489.
[38] B. Hinkov, A. Hugi, M. Beck, J. Faist, *Opt. Express* **2016**, *24*, 3294.
[39] E. Rodriguez, A. Mottaghizadeh, D. Gacemi, M. Jeannin, Z. Asghari, A. Vasanelli, Y. Todorov, Q. J. Wang, C. Sirtori, *Laser Photonics Rev.* **2020**, *14*, 1900389.
[40] D. Burghoff, Y. Yang, D. J. Hayton, J.-R. Gao, J. L. Reno, Q. Hu, *Opt. Express* **2015**, *23*, 1190.
[41] F. Cappelli, L. Consolino, G. Campo, I. Galli, D. Mazzotti, A. Campa, M. Siciliani de Cumis, P. Cancio Pastor, R. Eramo, M. Rösch, M. Beck, G. Scalari, J. Faist, P. De Natale, S. Bartalini, *Nat. Photonics* **2019**, *13*, 562.
[42] D. Oustinov, N. Jukam, R. Rungsawang, J. Madéo, S. Barbieri, P. Filloux, C. Sirtori, X. Marcadet, J. Tignon, S. Dhillon, *Nat. Commun.* **2010**, *1*, 69.
[43] S. Barbieri, M. Ravaro, P. Gellie, G. Santarelli, C. Manquest, C. Sirtori, S. P. Khanna, E. H. Linfield, A. G. Davies, *Nat. Photonics* **2011**, *5*, 306.

# Supplementary material of "RF Injection Locking of THz Metasurface Quantum-Cascade VECSEL"

Yu Wu,[1*] Christopher A. Curwen,[2] Mohammad Shahili,[1] John L. Reno,[3] Benjamin S. Williams[1]

## S1. Modeling of QC-metasurface and output coupler

The QC-metasurface shown in Figure 1 is modelled using full-wave 2D finite-element (FEM) simulation (Ansys HFSS), assuming it is infinite in extent, where a single metal-metal waveguide antenna is simulated with periodic boundary conditions. Simulated losses in the metal thin films are estimated using the Drude model ($n_{Au} = 5.9 \times 10^{22}$ cm$^{-3}$, $\tau_{Au,77\,K} = 39$ fs [1]), while for the semiconductor layer, band diagram is simulated to obtain the intersubband gain provided by the active material. The QC-active material is grown 10-μm thick by molecular beam epitaxy (wafer number VB0739). The active region is based upon a hybrid bound-to-continuum/ resonant-phonon design scheme and exhibits over 1 THz gain bandwidth peaking at 3.4 THz, the same one as used in Refs [2,3]. It consists of an GaAs/Al$_{0.15}$Ga$_{0.85}$As heterostructure where, starting from the injection barrier, the layer thicknesses in Å are **51**/103/**17**/107/**37**/88/**37**/172 (barrier layers are bold). The central 88 Å of the underlined well is Si-doped at $5 \times 10^{16}$ cm$^{-3}$. We simulated the band diagram for one module of the active region using a self-consistent Schrödinger-Poisson solver at the bias providing maximum gain (Figure S1(a)) and obtained its permittivity along the growth direction (Figure S1(b)). The dominant intersubband transition occurs between the upper lasing state 5 and the lower lasing state 4 at a frequency of $\nu_{54} = 3.4$ THz with oscillator strength of $f_{54} = 0.427$, where the population inversion takes up 25% of the total doping concentration. A good qualitative fitting of the permittivity is obtained by considering only the 5 → 4 transition using a Lorentz oscillator model (Figure S1(b), dashed lines):

$$\varepsilon_z(\omega) = \varepsilon_{core} + \frac{N_{ISB}e^2}{m^*L_{mod}}\frac{f_{54}}{\omega^2 - \omega_{54}^2 + i\omega\gamma}, \quad (1)$$

where $N_{ISB}$ is the population inversion sheet density per module, $\varepsilon_{core}$ is the semiconductor permittivity excluding free carrier contributions, $m^*$ is the GaAs electron effective mass, $L_{mod}$ is the length of one module of the active region and $\gamma$ is the damping term which is set as $2\pi \times 700$ GHz for best fit. The resonance frequency offset between the Lorentz model and Schrödinger simulation results can be explained by uncertainties in growth thicknesses and compositions of the QC-material. This permittivity is brought into FEM simulation, where the simulated reflectance and GDD are plotted in Figure 1(c).

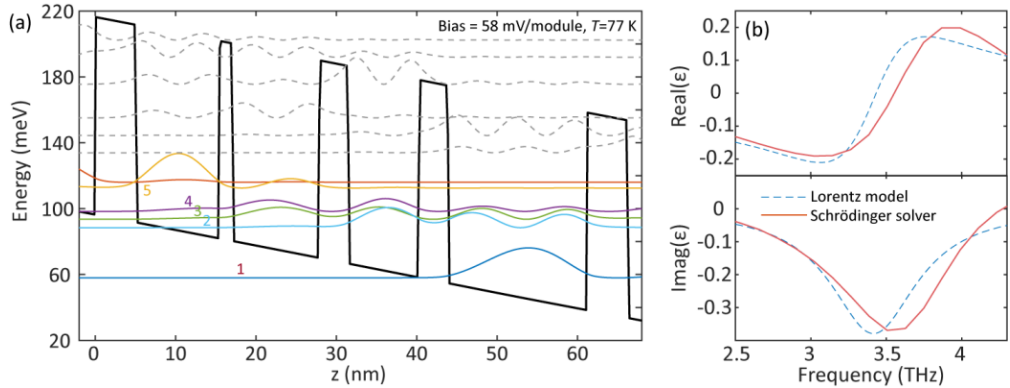

**Figure. S1.** (a) Conduction band diagram of the active region at the bias of 58 mV/module. (b) The real and imaginary part of the permittivity of the active region simulated using the Schrödinger-Poisson solver (red solid curves), and fitting based on Lorentz model (blue dashed curves).

The output coupler used to form a laser cavity is the same type that has been used in previous QC-VECSEL experiments [4,5]. It is made of an inductive Ti/Au mesh evaporated on a 100-µm-thick double-side-polished z-cut quartz substrate. The mesh is designed with a period of 13 µm and width of 3 µm, which determines the overall transmittance magnitude.

## S2. Effects of optical feedback

We experimentally found that optical feedback, even weak feedback originating from the FTIR mirrors, induces few-mode lasing in free-running QC-VECSELs. This can be useful, as it provides exact information on round-trip frequency from an observed beat note (Figure 1(e-f)) – without feedback, the device lases in single-mode regime with no beat note detected. It is believed that the performances (e.g. output power, spectral characteristics, injection locking range) of QC-lasers will be affected by the strength, length (phase) and the tilted angle of the optical feedback [6–9]. A systematic study will be needed using a motorized translational and rotational stage that can precisely control the optical feedback length and angle, and a rotatable polarizer that adjusts optical feedback strength.

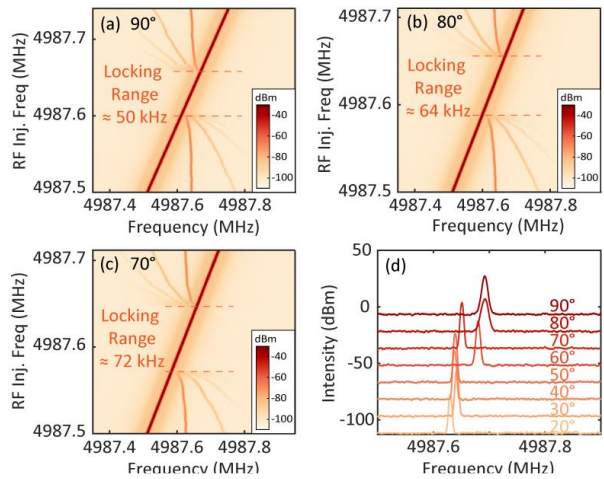

**Figure. S2.** (a-c) Beat note spectral maps under constant RF injection power of -15 dBm in the case when the strength of optical feedback is reduced controlled by the rotational angle from "90°" to "70°". (d) The free-running beat note frequency is shifted with reduced strength of optical feedback.

Here, we did a simple experiment to qualitatively demonstrate the effects of optical feedback strength on free-running beat note frequency as well as the RF injection locking range. We put a flat mirror in front of the cryostat window (approximately 15 cm from the device) and a rotatable wire-grid polarizer in between. THz radiation coming from the QC-VECSEL has a polarization perpendicular to the ridge antennas and the QC-metasurface only interacts with light at that polarization. We label it as "90°" for the case when 100% of the THz radiation passes through the polarizer. As the polarizer was rotated from "90°" to "70°", the amount of light feedbacked back into the QC-device was reduced. The collected beat note spectral maps under a constant RF power of -15 dBm are plotted in Figure S2(a-c) which indicate an increasing locking range with respect to the reduced optical feedback strength. Moreover, the free-running beat notes were also collected in Figure S2(d) showing frequency shift as optical feedback strength was changed.

## S3. Transmission loss in the QC-device

To estimate the transmission loss in the QC-device due to impedance mismatch, an FEM simulation is used accounting for the finite dimension of the metasurface including its electrical packaging. Figure S3(a) shows the QC-device mounted in the focusing cavity that is modelled using Ansys HFSS in Figure S3(b). Only the electrically biased ridges are modelled with metal layers loaded with Drude loss; the circular biased area is assumed to be loaded with GaAs with a shunt conductivity derived from experimental $dI/dV$ curve while the unbiased area is defined as bulk GaAs. Two 1-mil bond wires of approximately 2 mm length electrically connect the metasurface to a "gold pad": a 2.5 mm × 4.5 mm × 0.254 mm thick $Al_2O_3$ pad coated with Au above and below. The pad is soldered to the center pin of an SMA connector on the other side. The E-field distribution along one of the biased ridges is simulated at 5 GHz which indicates the injected RF signal propagating along the QC-metasurface has an effective wavelength much longer than its dimension. The simulated power dissipation with a 50 Ω excitation port is plotted in Figure S3(c), which gives an estimated power transfer of 4% from the SMA into the QC-device and shows good agreement with circuit model result (see section S4) except at lower frequencies. The frequency of the electrical resonance is determined by the dimensions of the gold pad and bond wires.

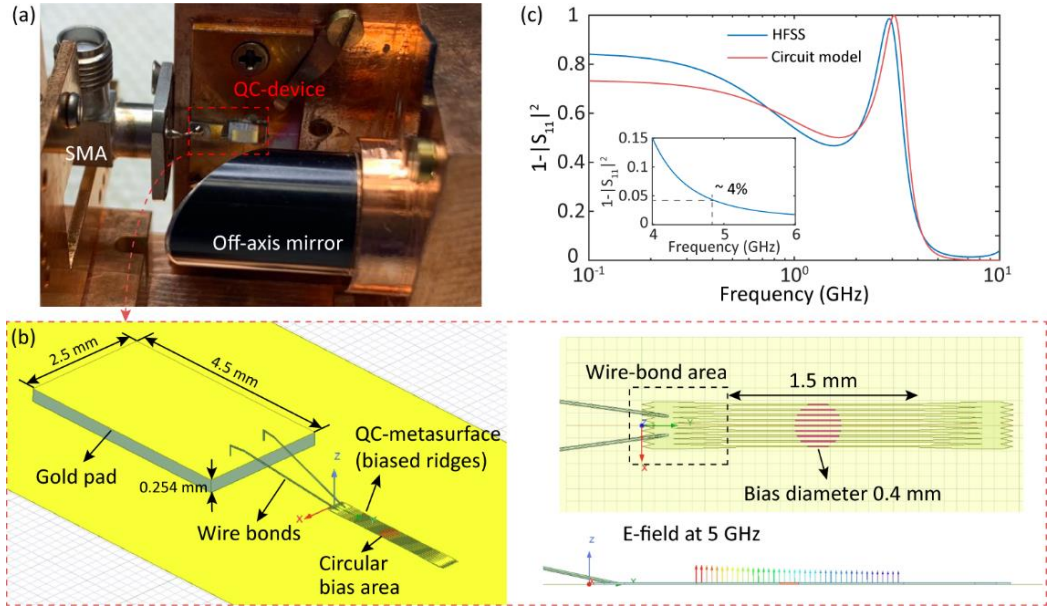

**Figure. S3.** (a) Image of QC metasurface device mounted in the focusing cavity. (b) Electrical packaging structure that is modelled in Ansys HFSS including the gold pad and bond wires. Their dimensions are labeled and the simulated E-field distribution along one of the biased ridges at 5 GHz is plotted. (c) Transmission coefficient simulated within HFSS assuming an excitation port of 50 Ω (blue) and calculated using the lumped element circuit model (red).

## S4. Microwave rectification measurement and equivalent lumped element circuit model

To experimentally characterize the response of QC-device to injected RF signal, microwave rectification technique has been applied as described in Ref. [10–12]. The RF signal generated from the synthesizer is amplitude modulated at a frequency of 10 kHz and injected into the QC-device, while the latter is biased at a constant current. The variation in DC rectification voltage is measured using a lock-in amplifier referenced to the amplitude modulation. The injected RF power is kept constant at -10 dBm, and the normalized rectification voltage (proportional to $|V_{RF}|^2$) is plotted in Figure S4 at bias current of 0.235 mA. The result is well described by a lumped-element circuit model described in more details below. Notably, an electrical resonance is present at 3 GHz (associated with the LC parasitics) and the 3-dB cutoff frequency is 3.7 GHz followed by a rapid roll-off at higher frequencies. Consequently, the rectification voltage of the QC-device at the target frequency of 4.8 GHz is reduced to ~5% of that at lower frequencies.

An equivalent lumped-element circuit is introduced to model the QC-device and explain the rectification measurement. The QC-metasurface used in this paper has a dimension much smaller than the RF operation wavelength as shown in Figure S3(b) and is therefore represented as a parallel plate capacitor in parallel with a resistor coming from the effect of QC-active material. The capacitance $C_{MS}$ is calculated based on the dimension and thickness of the metasurface assuming the permittivity of 12.5 for the GaAs/AlGaAs active region. The differential shunt resistance $R_{AR}$ is obtained based on the experimental slope of *I-V* curve at the bias point. $L_{wire}$ is the inductance of the two wire bonds connected in series with the RC circuit and is estimated based upon the rule of thumb of ~1nH/mm/wire. $C_{gold\ pad}$ is the capacitance of the gold pad whose value is estimated based on its dimension and the permittivity of $Al_2O_3$. The gold pad is connected to the RF source with a 50Ω generator impedance $R_g$, which provides a RF power of $P_{RF}$ and an equivalent RF voltage of $V_{RF} = 2\sqrt{R_g P_{RF}}$.

The rectification voltage of the QC-device can be calculated according to Ref. [11]:

$$V_{rect} = \frac{1}{2}|V''|_{I_0} I_{RF,QCL}^2, \quad (S1)$$

where $V''$ is the second derivative of the *I-V* curve at DC bias current $I_0$. $I_{RF,\ QCL}$ is the RF modulation current injected into the QC-active material:

$$I_{RF,QCL} = V_{RF} \frac{R_L}{R_{AR}(Z_{QCL} + R_L)} \frac{Z_{MS}}{(j\omega L + Z_{MS})}, \quad (S2)$$

where the impedance of the QC-device $Z_{QCL}$ and the metasurface $Z_{MS}$ is pointed out in Figure S4 inset. The theoretical rectification voltage is plotted in dashed line using values of: $C_{MS}$ = 6.1 pF, $R_{AR}$ = 16 Ω, $C_{gold\ pad}$ = 3.8 pF, $L_{wire}$ = 1nH.

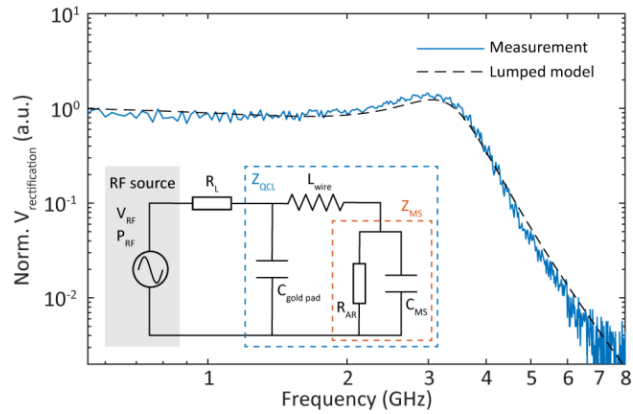

**Figure. S4.** Normalized rectification curves (solid line) measured at bias current of 0.235 mA, together with the theoretical fits (dashed line) obtained based on lumped element circuit model. Inset: equivalent lumped element circuit model. Blue and red dashed boxes point out the impedances of the QC-device and metasurface.

References:
[1] N. Laman, D. Grischkowsky, *Appl. Phys. Lett.* **2008**, *93*, 051105.
[2] C. A. Curwen, J. L. Reno, B. S. Williams, *Nat. Photonics* **2019**, *13*, 855.
[3] C. A. Curwen, J. L. Reno, B. S. Williams, *Electron. Lett.* **2020**, *56*, 1264.

[4]   L. Xu, C. A. Curwen, J. L. Reno, B. S. Williams, *Appl. Phys. Lett.* **2017**, *111*, 101101.
[5]   Y. Wu, S. Addamane, J. L. Reno, B. S. Williams, *Appl. Phys. Lett.* **2021**, *119*, 111103.
[6]   M. Wienold, B. Röben, L. Schrottke, H. T. Grahn, *Opt. Express* **2014**, *22*, 30410.
[7]   X. Liao, X. Wang, K. Zhou, W. Guan, Z. Li, X. Ma, C. Wang, J. C. Cao, C. Wang, H. Li, *Opt. Express* **2022**, *30*, 35937.
[8]   D. S. Seo, J. D. Park, J. G. Mcinerney, M. Osinski, *IEEE J. Quantum Electron.* **1989**, *25*, 2229.
[9]   X. G. Wang, B. Bin Zhao, Y. Deng, V. Kovanis, C. Wang, *Phys. Rev. A* **2021**, *103*, 23528.
[10]  E. Rodriguez, A. Mottaghizadeh, D. Gacemi, M. Jeannin, Z. Asghari, A. Vasanelli, Y. Todorov, Q. J. Wang, C. Sirtori, *Laser Photonics Rev.* **2020**, *14*, 1900389.
[11]  B. Hinkov, A. Hugi, M. Beck, J. Faist, *Opt. Express* **2016**, *24*, 3294.
[12]  L. Gu, W. J. Wan, Y. H. Zhu, Z. L. Fu, H. Li, J. C. Cao, *J. Opt. (United Kingdom)* **2017**, *19*, 065706.